\documentclass[twocolumn,journal]{IEEEtran}
\usepackage[T1]{fontenc}
\usepackage[latin9]{inputenc}
\usepackage{amsmath}
\usepackage{amssymb}
\usepackage{graphicx}
\usepackage[unicode=true,
 bookmarks=true,bookmarksnumbered=true,bookmarksopen=true,bookmarksopenlevel=1,
 breaklinks=false,pdfborder={0 0 0},pdfborderstyle={},backref=false,colorlinks=false]
 {hyperref}
\hypersetup{pdftitle={Your Title},
 pdfauthor={Your Name},
 pdfborderstyle={},pdfborderstyle={},pdfborderstyle={},pdfborderstyle={},pdfpagelayout=OneColumn,pdfnewwindow=true,pdfstartview=XYZ,plainpages=false}

\makeatletter

\providecommand{\tabularnewline}{\\}

\usepackage[noadjust]{cite}

\makeatother

\begin{document}
\title{Deep Energy Autoencoder for Noncoherent Multicarrier MU-SIMO Systems}
\author{Thien Van Luong, Youngwook Ko,~\IEEEmembership{Senior Member,~IEEE,}
Ngo Anh Vien, \\
 Michail Matthaiou,~\IEEEmembership{Senior Member,~IEEE,} and Hien
Quoc Ngo,~\IEEEmembership{Member,~IEEE} \thanks{This work was supported by the Engineering and Physical Sciences Research
Council under Grant EP/N509541/1. The work of M. Matthaiou was supported
by the RAEng/The Leverhulme Trust Senior Research Fellowship under
Grant LTSRF1718\textbackslash 14\textbackslash 2 and by a research
grant from the Department for the Economy Northern Ireland under the
US-Ireland R\&D Partnership Programme. The work of H. Q. Ngo was supported
by the UK Research and Innovation Future Leaders Fellowships under
Grant MR/S017666/1.}\thanks{Thien Van Luong, Ngo Anh Vien, Michail Matthaiou, and Hien Quoc Ngo
are with the ECIT Institute, Queen's University Belfast, Belfast,
BT3 9DT, UK, (email: \{tluong01, v.ngo, m.matthaiou, hien.ngo\}@qub.ac.uk). }\thanks{Youngwook Ko is with the University of York, Heslington, York, YO10
5DD, UK, (email: youngwook.ko@york.ac.uk).}}
\maketitle
\begin{abstract}
We propose a novel deep energy autoencoder (EA) for noncoherent multicarrier
multiuser single-input multiple-output (MU-SIMO) systems under fading
channels. In particular, a single-user noncoherent EA-based (NC-EA)
system, based on the multicarrier SIMO framework, is first proposed,
where both the transmitter and receiver are represented by deep neural
networks (DNNs), known as the encoder and decoder of an EA. Unlike
existing systems, the decoder of the NC-EA is fed only with the energy
combined from all receive antennas, while its encoder outputs a real-valued
vector whose elements stand for the sub-carrier power levels. Using
the NC-EA, we then develop two novel DNN structures for both uplink
and downlink NC-EA multiple access (NC-EAMA) schemes, based on the
multicarrier MU-SIMO framework. Note that NC-EAMA allows multiple
users to share the same sub-carriers, thus enables to achieve higher
performance gains than noncoherent orthogonal counterparts. By properly
training, the proposed NC-EA and NC-EAMA can efficiently recover the
transmitted data without any channel state information estimation.
Simulation results clearly show the superiority of our schemes in
terms of reliability, flexibility and complexity over baseline schemes. 
\end{abstract}


\section{Introduction}

Multicarrier transmission has become a key technology for numerous
wireless systems due to its simple implementation and robustness against
inter-symbol interference and delay spreading caused by multipath
fading. Orthogonal frequency division multiplexing (OFDM) \cite{OFDMsurvey},
which is the most popular multicarrier technique, has been included
in various standards, such as Wi-Fi 802.11 and 3GPP's LTE. In general,
OFDM is not only spectrally efficient, but also enables the use of
low-complexity transceivers as it only needs one-tap equalizer per
sub-carrier to effectively combat multipath fading effects.

Over the past years, many efforts have been made to improve the reliability
and spectral efficiency (SE) of multicarrier systems; in this context,
OFDM with index modulation (OFDM-IM) \cite{basar3013} has recently
emerged as a promising technique to replace conventional OFDM. In
particular, OFDM-IM activates only a subset of sub-carriers to carry
additional data bits via the indices of active sub-carriers without
any extra needs of bandwidth or power. The error performance of OFDM-IM
under channel state information (CSI) uncertainty was comprehensively
analyzed in \cite{ThienTVT2017,thienBERGD} with the maximum likelihood
(ML) and energy-based greedy (GD) detectors \cite{GDjamesPIMRC2015}.
The reliability of OFDM-IM can be further enhanced by using coordinate
interleaving \cite{CIbasar2015}, repetition \cite{ThienTWC2018,codedIM2017choi}
and spreading codes \cite{ThienTVT2018}, while its SE can be increased
by relaxing the number of active sub-carriers \cite{GeneralizedIM}.
Note that the aforementioned multicarrier schemes are based on coherent
detection designs, where the receiver needs to estimate the CSI of
all sub-carriers regardless of their activity. As a result, they may
suffer from a high pilot signaling overhead, particularly under fast-varying
fading channels. Thus, in \cite{NoncoherentOFDMIM}, noncoherent OFDM-IM
(NC-OFDM-IM), also known as a generalized version of frequency shift
keying (FSK), was introduced, which uses only the active indices to
convey data bits. In fact, this scheme utilizes simple unitary codewords,
i.e., transmitted vectors, as it allows a fixed number of active sub-carriers
to carry the same power. Yet, this design may not be optimal, especially
when some of combinations of active indices are redundant. We aim
to address this issue by devising an optimal codeword design for a
noncoherent multicarrier (NC-MC) energy-based scheme, using deep learning
(DL) tools \cite{SCHMIDHUBER201585}.

Regarding noncoherent single-carrier transmissions, various energy-based
detection (ED) schemes with nonnegative pulse amplitude modulation
(PAM) have been investigated, especially in massive single-input multiple-output
(SIMO) systems. For example, the performance of an ED-based massive
SIMO system was analyzed in \cite{Ham2015}, which results in an optimal
power allocation design. In \cite{Xie2019}, the effects of correlated
Rayleigh fading on the performance of a similar system were investigated.
In \cite{PamMed2016}, the PAM constellations that maximize the minimum
Euclidean distance (MED) between signal points were designed, where
the resulting scheme can be termed as PAM-MED. This work also looked
into the constellation design for two users, which iteratively uses
two separate constellations in two time slots. In \cite{Leung2017},
a uniquely factorable hexagonal constellation was proposed for noncoherent
SIMO systems, where the channels are assumed to remain unchanged in
each two time slots. Besides, the constellation designs under different
assumptions of the CSI statistics were presented in \cite{Alex2016},
while the channel gains were used for optimizing the PAM constellations
in \cite{Mallik2018}. We note that most of existing works have addressed
the noncoherent ED-based single-carrier and single-user systems, while
the NC-MC energy-based designs for multiuser SIMO (MU-SIMO) transmissions
have been overlooked. Our work aims to fill this fundamental gap.

Recently, DL based on deep neural networks (DNNs) \cite{SCHMIDHUBER201585}
has emerged as a powerful tool to address diverse problems in physical-layer
wireless communications. For instance, in \cite{YeOFDM2018}, channel
estimation and signal detection of OFDM systems were performed by
DNNs, while in \cite{DeepIM2019} a DL-based detector, called as DeepIM,
was proposed for OFDM-IM. Particularly, in \cite{Oshea2017}, a novel
end-to-end learning-based system was proposed, where both the transmitter
and receiver are represented by DNNs, which are known as the encoder
and decoder of an AE. This data-driven system enables a joint optimization
of both the transmitter and receiver via training, leading to better
performance than conventional block-based systems. The AE-based system
was implemented under real-world environments in \cite{Dorner2018}.
The AE concept was also applied to OFDM and noncoherent MU-SIMO systems
in \cite{AEofdm2018} and \cite{Xue2019}, respectively. Some end-to-end
AE-based schemes under unknown channel models were proposed in \cite{Aou2018,YeGan2018},
which aim to eliminate the need of a differentiable channel model.
Note that under fading channels, these learning-based schemes have
to employ pilot transmissions to estimate the CSI for signal detection.
To the best of our knowledge, none of the existing works has explored
the potential of DL in the noncoherent ED-based systems.

In this paper, DL is first applied to noncoherent energy-based systems
to improve the performance over current ED systems. Our main contributions
are summarized as follows: 
\begin{itemize}
\item We propose a novel deep energy autoencoder (EA) for single-user multicarrier
SIMO systems, coined as NC-EA, whose transmitter and receiver are
modeled as the encoder and decoder (DNNs) of an EA. Unlike existing
schemes \cite{Oshea2017} which utilize complex signals, the encoder
of NC-EA outputs a real-valued vector whose elements represent the
sub-carrier power levels, while its decoder is fed only with the combined
energy of signals from the receive antennas without any knowledge
of CSI. 
\item Using NC-EA, we construct two novel DNN structures for both downlink
and uplink NC-EA multiple access (NC-EAMA) schemes, in the multicarrier
MU-SIMO framework. Note that NC-EAMA allows multiple users to access
the same set of sub-carriers, thus can be considered as a type of
noncoherent non-orthogonal multiple access (NC-NOMA), which is expected
to achieve higher performance gains than the noncoherent energy-based
orthogonal schemes, termed as NC-OMA. 
\item Various simulations clearly present that by properly training with
simulated data, the proposed learning-based schemes can efficiently
decode data without any CSI estimation, and outperform the hand-crafted
baselines at reduced complexity. In this context, our schemes are
very attractive for various machine-type communications (MTCs) \cite{mMTC}
which require reliable, low latency and low complexity connectivity. 
\end{itemize}
The rest of the paper is as follows: Section II presents the single-user
NC-EA, while Section III introduces the uplink and downlink NC-EAMA
systems. Simulation results are provided in Section IV. Finally, Section
V concludes the paper.

\textit{Notation:} Upper-case bold and lower-case bold letters present
matrices and vectors, respectively; $C(n,k)$ denotes the binomial
coefficient for $n$ choose $k$; $\left\lfloor .\right\rfloor $
denotes the floor function; $(.)^{T}$ and $\left\Vert .\right\Vert $
stand for the transpose operation and the Frobenius norm, respectively.
$\mathcal{CN}\left(0,\sigma^{2}\right)$ denotes the complex Gaussian
distribution with zero mean and variance $\sigma^{2}.$

\section{Single-User NC-EA System}

\subsection{NC-EA Structure}

\begin{figure}[tb]
\begin{centering}
\includegraphics[width=0.9\columnwidth]{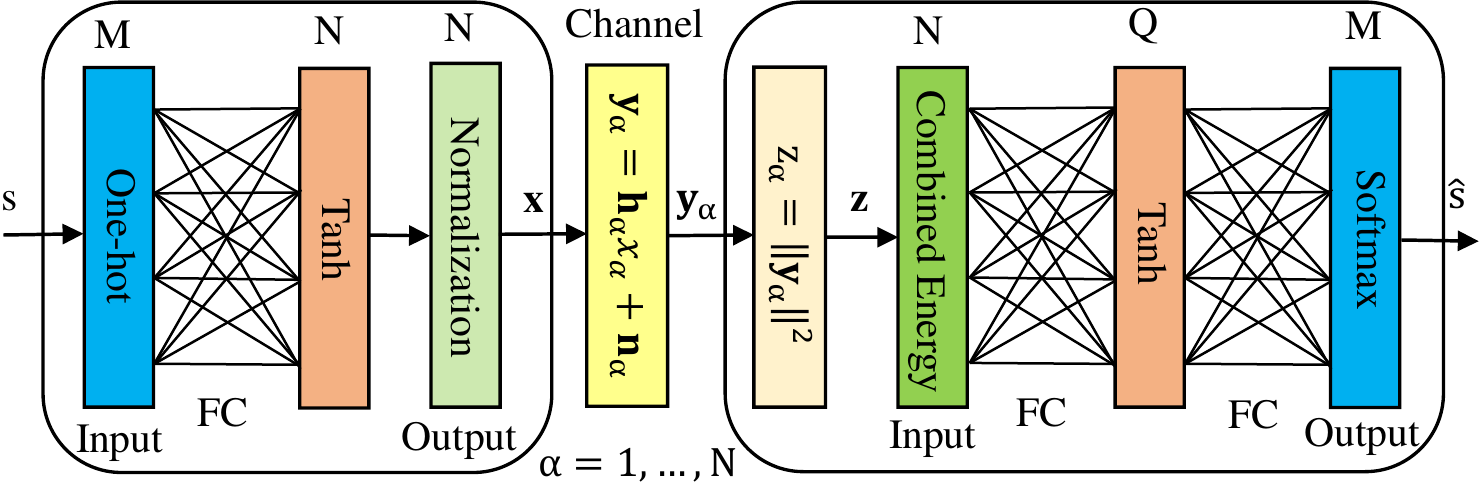} 
\par\end{centering}
\caption{Structure of the single-user NC-EA system. \label{fig:EAE-struc}}
\end{figure}

Consider a NC-MC SIMO system with $N$ sub-carriers, which does not
require any CSI estimation at the transmitter and the receiver. We
assume that the transmitter has a single antenna, while the receiver
has $L$ antennas. Unlike current NC-MC schemes, such as NC-OFDM-IM
\cite{NoncoherentOFDMIM}, we implement both the transmitter and receiver
of NC-MC by DNNs, proposing a deep energy autoencoder (EA) structure
in Fig. \ref{fig:EAE-struc}, where the resulting scheme can be termed
as NC-EA. Note that the proposed EA differs from the conventional
AE \cite{Oshea2017} in that the decoder of the EA is fed only with
the combined energy from the receive antennas, without any knowledge
of CSI.

In particular, the NC-EA structure consists of the encoder and decoder
neural networks, which represent the transmitter and receiver, respectively.
At the transmitter, the incoming message $s\in\mathcal{S}=\left\{ s_{1},...,s_{M}\right\} $
is mapped to an $M\times1$ one-hot vector, which is used as an input
vector of the encoder, wherein $\mathcal{S}$ is the set of all $M=2^{m}$
possible messages, each having $m$ data bits. Note that the one-hot
vector $\mathbf{s}$ has a single unit entry which is indexed by $s$
in $\mathcal{S}$, while the remaining entries are zeros. The encoder
has a full-connected (FC) layer with the hyperbolic tangent (Tanh)
activation function \cite{SCHMIDHUBER201585}, whose output is given
by $\mathbf{u}=\sigma_{\text{Tanh}}\left(\mathbf{W}\mathbf{s}+\mathbf{b}\right)$,
where $\mathbf{W}$ and $\mathbf{b}$ are the $N\times M$ weight
matrix and $N\times1$ bias vector, respectively, and $\sigma_{\text{Tanh}}$
denotes the element-wise Tanh function. Then, $\mathbf{u}$ is normalized
to constrain the average transmit power over each sub-carrier to be
a given constant, as follows:

\begin{equation}
\mathbf{x}=\frac{\sqrt{NSE_{s}}\mathbf{u}}{\sqrt{\sum_{i=1}^{S}\left\Vert \mathbf{u}_{i}\right\Vert ^{2}}},\label{eq:norm}
\end{equation}
where $E_{s}$ is the average transmit power per sub-carrier and $\mathbf{u}_{i}=\sigma_{\text{Tanh}}\left(\mathbf{W}\mathbf{s}_{i}+\mathbf{b}\right)$
with $\mathbf{s}_{i}\in\Omega=\left\{ \mathbf{s}_{1},...,\mathbf{s}_{T}\right\} $
which is a batch of $S$ training samples ($T$ is coined as the batch
size). The set of all possible $M$ codewords $\mathbf{x}$ can be
considered as a codebook of NC-EA denoted by $\mathcal{X}=\left\{ \mathbf{x}_{1},...,\mathbf{x}_{M}\right\} $,
while the mapping from $\mathbf{s}$ to $\mathbf{x}$ can be represented
by $\mathbf{x}=f_{\theta_{\text{enc}}}\left(\mathbf{s}\right)$, where
$\theta_{\text{enc}}=\left\{ \mathbf{W},\mathbf{b}\right\} $ denotes
the parameters of the encoder.

Note that the average power normalization over an entire training
batch in \eqref{eq:norm} is preferable for the energy detection of
an EA than a fixed power constraint with $\mathbf{x}=\sqrt{NE_{s}}\mathbf{u}/\left\Vert \mathbf{u}\right\Vert $,
in order to make the codewords' energies as different as possible.
It is also worth noting that the codeword $\mathbf{x}$ is a real-valued
vector whose entries indicate the amplitudes of sub-carrier symbols;
hence, it is suitable for an energy-based decoder of NC-EA. This real-valued
design also reduces the model complexity of NC-EA, which facilitates
the training to converge faster, compared to the complex-valued design
\cite{Oshea2017}.

The received signal vector from $L$ receive antennas, in frequency
sub-carrier $\alpha$, for $\alpha=1,...,N$, is given by 
\begin{equation}
\mathbf{y}_{\alpha}=\mathbf{h}_{\alpha}x_{\alpha}+\mathbf{n}_{\alpha},\label{eq:y_a}
\end{equation}
where $\mathbf{y}_{\alpha}=\left[y_{1}\left(\alpha\right),....,y_{L}\left(\alpha\right)\right]^{T}$,
$x_{\alpha}$ is the $\alpha$-th entry of $\mathbf{x}$, $\mathbf{h}_{\alpha}=\left[h_{1}\left(\alpha\right),...,h_{L}\left(\alpha\right)\right]^{T}$
denotes the Rayleigh fading channel vector from the transmitter to
$L$ receive antennas with $h_{l}\left(\alpha\right)\sim\mathcal{CN}\left(0,1\right)$,
and $\mathbf{n}_{\alpha}$ is the additive noise vector with $n_{l}\left(\alpha\right)\sim\mathcal{CN}\left(0,N_{0}\right)$,
for $l=1,...,L$. We assume that the entries of $\mathbf{h}_{\alpha}$
and $\mathbf{n}_{\alpha}$ are independent and identically distributed
(i.i.d.) random variables (RVs). Hence, the average signal-to-noise
ratio (SNR) per sub-carrier is $\bar{\gamma}=E_{s}/N_{0}$.

As for the NC-EA decoder, the combined energy received from $L$ receive
antennas is first computed for each sub-carrier:

\begin{equation}
z_{\alpha}=\left\Vert \mathbf{y}_{\alpha}\right\Vert ^{2}=\sum_{l=1}^{L}\left|y_{l}\left(\alpha\right)\right|^{2},\label{eq:z_a}
\end{equation}
which produces the $N\times1$ combined energy vector for all sub-carriers
$\mathbf{z}=\left[z_{1},...,z_{N}\right]^{T}$. This received energy
vector is then fed to the DNN of the decoder as shown in Fig. \ref{fig:EAE-struc}.
In particular, the proposed decoder structure has two non-linear FC
layers, in which the first FC layer has $Q$ nodes with the Tanh activation,
while the second FC layer is the output layer of $M$ nodes with the
softmax activation \cite{SCHMIDHUBER201585}.\footnote{The number of hidden layers is miminized based on experiments in order
to make the DNN model of NC-EA perform best at a reduced complexity.
Moreover, we use Tanh at both the encoder and decoder of NC-EA since
this activation always offers better performance than others such
as Linear, Sigmoid and Relu activations \cite{SCHMIDHUBER201585},
as observed through our experiments.} Let $\theta_{\text{dec}}=\left\{ \mathbf{W}_{i},\mathbf{b}_{i}\right\} _{i=1,2}$
be the weights and biases of the two decoder layers. The output vector
of the softmax layer is mathematically expressed by 
\begin{equation}
\hat{\mathbf{s}}=f_{\theta_{\text{dec}}}\left(\mathbf{z}\right)=\sigma_{\text{Softmax}}\left(\mathbf{W}_{2}\sigma_{\text{Tanh}}\left(\mathbf{W}_{1}\mathbf{z}+\mathbf{b}_{1}\right)+\mathbf{b}_{2}\right),\label{eq:s_hat}
\end{equation}
where $\sigma_{\text{Softmax}}$ denotes the element-wise softmax
function. The estimated $\hat{s}$ is determined based on the largest
element of $\hat{\mathbf{s}}$.

We note that since the DNN decoder obtained via training may not be
optimal for certain values of $N$ and $M$, the optimal noncoherent
ML may be used to improve the performance of NC-EA compared to using
the DNN decoder, as follows: 
\begin{equation}
\hat{\mathbf{x}}=\arg\min_{\mathbf{x}\in\mathcal{X}}\sum_{\alpha=1}^{N}\left[\frac{z_{\alpha}}{\left|x_{\alpha}\right|^{2}+N_{0}}+L\ln\left(\left|x_{\alpha}\right|^{2}+N_{0}\right)\right],\label{eq:ML}
\end{equation}
where we have followed the derivation of \cite[Chapter 5]{DetectTheory}.

It is worth noting that the NC-EA requires only the received energy
for signal decoding, thus does not involve any channel estimation,
which is particularly desirable for low latency and complexity communications.
More importantly, our scheme provides a number of advantages over
existing hand-designed schemes, such as NC-OFDM-IM \cite{NoncoherentOFDMIM},
as follows: 
\begin{itemize}
\item The NC-EA can send any number of data bits $m$ for given $N$, while
that of NC-OFDM-IM is limited to $m_{0}=\left\lfloor \log_{2}C\left(N,K\right)\right\rfloor $
bits, where $K$ is the number of active sub-carriers. Hence, our
scheme is not only more flexible but also is able to support higher
data rates than its counterpart. For example, when $N=4$, NC-OFDM-IM
supports only $m_{0}\le2$ bits for every $K<N$, while NC-EA supports
more bits with $m=3$ or even $4$ bits. 
\item The NC-EA can achieve higher reliability than NC-OFDM-IM since our
scheme benefits from a joint optimization of both the transmitter
and the receiver through training the EA model to achieve an optimal
design of codewords $\mathcal{X}$. 
\item The decoder of an NC-EA-based system is very simple with only one
hidden layer of $Q$ nodes. Hence, when $Q$ is not too large, NC-EA
can achieve even lower decoding complexity than NC-OFDM-IM. This will
be verified in Subsection IV.B. 
\item The NC-EA concept can be extended to NC-NOMA, where multiple users
share the same $N$ sub-carriers for the NC-EA transmission, as will
be shown in Section III. Note that this important benefit is not available
in current NC-MC schemes, whose hand-designed energy-based detector
is only applicable to single-user schemes. 
\end{itemize}
In summary, apart from no channel estimation, the proposed NC-EA achieves
higher flexibility and reliability at even lower computational complexity
than existing schemes. Hence, NC-EA can be easily implemented in small
and low-cost devices such as sensors. These benefits make our scheme
attractive to various MTC applications \cite{mMTC} which demand reliable,
ultra-low latency and low-complexity connectivity.

\subsection{Training procedure of NC-EA}

The proposed NC-EA model is trained to minimize the difference between
the original vector $\mathbf{s}$ and its prediction $\hat{\mathbf{s}}$,
using dataset collected from simulations. More precisely, the training
dataset includes $\mathbf{s},$ $\mathbf{h}_{\alpha}$ and $\mathbf{n}_{\alpha}$
$\left(\alpha=1,...,N\right)$, in which the input one-hot vector
$\mathbf{s}$ is randomly generated and fed to the encoder, then the
channel and noise vectors $\mathbf{h}_{\alpha}$, $\mathbf{n}_{\alpha}$
are randomly generated and added to the output of the encoder. Then,
the output of the channel layer $\mathbf{y}_{\alpha}$ in \eqref{eq:y_a}
is used for the computation of the combined energy $\mathbf{z}$ for
each sub-carrier, which is fed to the DNN decoder from which $\mathbf{s}$
is recovered. We adopt the conventional mean squared error (MSE) loss
function for training the NC-EA as follows: 
\begin{equation}
\mathcal{L}\left(\theta\right)=\frac{1}{T}\sum_{i=1}^{T}\left\Vert \mathbf{s}_{i}-\hat{\mathbf{s}}_{i}\right\Vert ^{2},\label{eq:loss-NC-EA}
\end{equation}
where $\theta=\left\{ \theta_{\text{enc}},\theta_{\text{dec}}\right\} $
denotes the model parameters of NC-EA and $T$ is the training batch
size.\footnote{Note that based on our experiments, the MSE loss always offers comparable
or better performance than the cross-entropy loss, and thus in this
work we use the MSE loss only for training the proposed EA-based schemes.} Using \eqref{eq:loss-NC-EA}, the NC-EA model parameters are updated
based on the stochastic gradient descent (SGD) algorithm as follows:
\begin{equation}
\theta:=\theta-\eta\nabla\mathcal{L}\left(\theta\right),\label{eq:theta}
\end{equation}
where $\eta$ is the learning rate which regulates how much to adjust
the parameters. In this work, we adopt an advanced SGD-based update
method, named as adaptive moment estimation (Adam), along with the
Xavier method for initializations of weights and biases. Note that
these methods are available on various off-the-shelf DL libraries,
such as Tensorflow \cite{tensorflow2015}.

Since the NC-EA only utilizes the received energy for signal detection,
its decoding performance is highly sensitive to the SNR level $\bar{\gamma}$
used for training. This means that the NC-EA model trained with a
training SNR (denoted by $\bar{\gamma}_{\text{tr}}$) performs best
only at the testing SNRs (denoted by $\bar{\gamma}_{\text{te}}$)
that are close to $\bar{\gamma}_{\text{tr}}$, while it does not perform
well at other testing SNRs that are far from $\bar{\gamma}_{\text{tr}}$.
Hence, to overcome such overfitting problem, we train the NC-EA with
multiple SNRs and then test the trained models with $\bar{\gamma}_{\text{te}}=\bar{\gamma}_{\text{tr}}$.
As such, under varying channel variances, we need to retrain the model
once $\bar{\gamma}$ changes, or store multiple pre-trained models
with different $\bar{\gamma}_{\text{tr}}$. It is also necessary to
accurately choose the encoder-decoder pair corresponding to each SNR
before transmission. In order to reduce the training time when $L$
is very large, i.e., massive SIMO systems, we can train NC-EA with
small $L,$ using the average received energy $\bar{z}_{\alpha}=\left\Vert \mathbf{y}_{\alpha}\right\Vert ^{2}/L$
rather than $z_{\alpha}$ in \eqref{eq:z_a}, then the trained model
still works well for larger $L$.

Note that there are still some issues regarding training NC-EA in
practice. For example, in actual systems, the channel model and statistics
as described in \eqref{eq:y_a} may be completely unknown, and this
obviously hinders the channel gradient computation to update the transmitter.
To overcome this issue, reinforcement learning \cite{Aou2018} or
generative adversarial networks \cite{YeGan2018}, which has recently
been used to learn the channel model of end-to-end learning-based
communication systems, can be applied to NC-EA. Yet, such extensions
are far beyond the scope of this work and will be part of our future
work.

\section{NC-EA Multiple Access Systems}

Using the NC-EA, we propose two novel DNN structures for both uplink
and downlink NC-EA multiple access (NC-EAMA) systems also based on
multicarrier SIMO framework. Note that the proposed NC-EAMA is able
to allow multiple users to share the same set of frequency sub-carriers.
Then, to improve the performance, a new loss function for training
NC-EAMA is designed, which ensures not only fast training convergence
but also fairness performance among users.

\subsection{Uplink NC-EAMA}

\begin{figure}[tb]
\begin{centering}
\includegraphics[width=1\columnwidth]{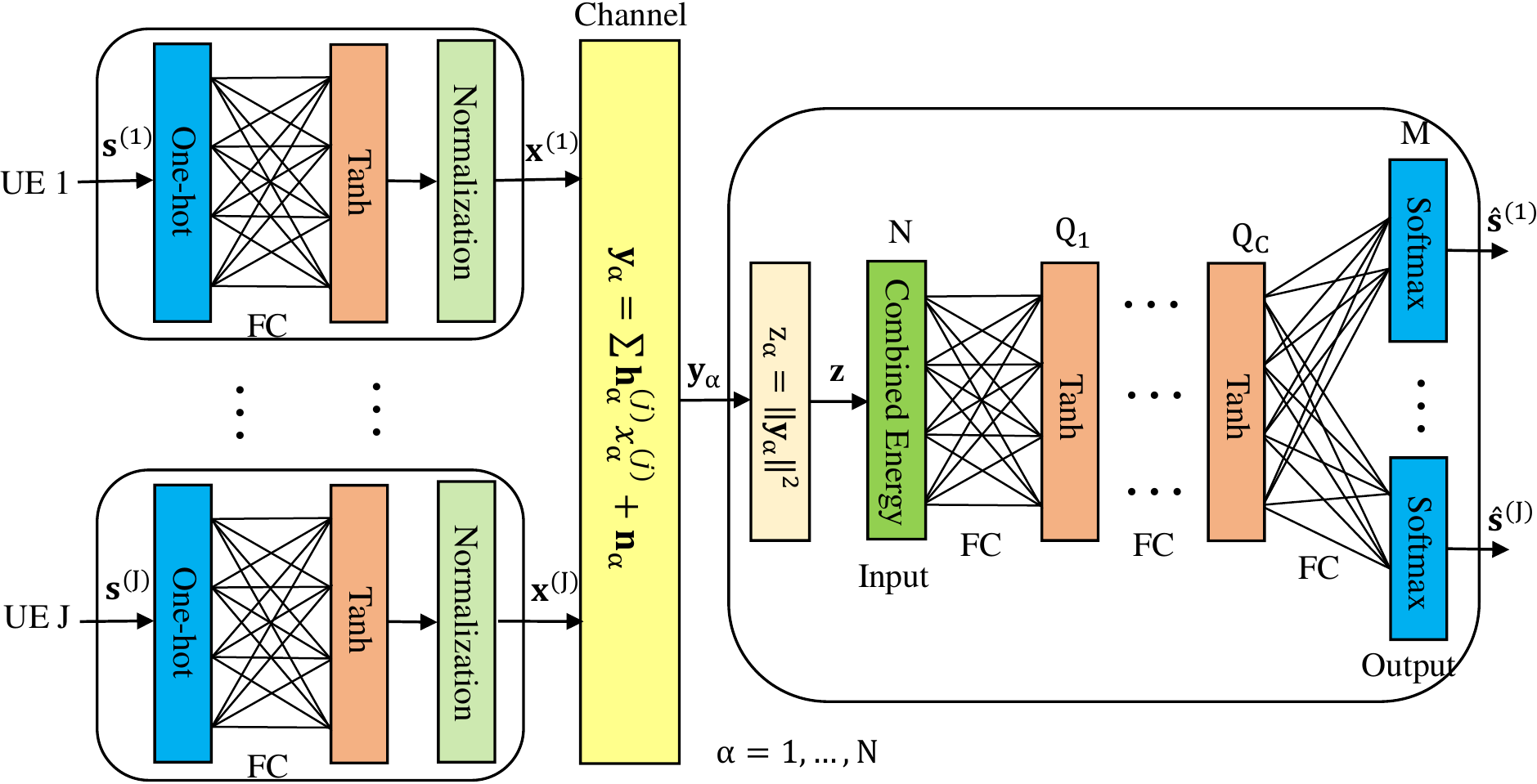} 
\par\end{centering}
\caption{Structure of the uplink NC-EAMA system. \label{fig:EAMA-UL-struc}}
\end{figure}

The proposed structure of uplink NC-EAMA is depicted in Fig. \ref{fig:EAMA-UL-struc},
where $J$ single-antenna users simultaneously send their data to
a central access point (AP) equipped with $L$ antennas, using the
same $N$ sub-carriers for NC-EA transmissions. Like NC-EA, the AP
in uplink NC-EAMA does not require any CSI knowledge of users in the
detection process. In particular, all users employ the same encoder
structure as that of single-user NC-EA, which makes them have the
same average transmit power. We assume that $\mathbf{s}^{\left(j\right)}$
is the input vector of user $j$'s encoder, while $\mathbf{x}^{\left(j\right)}$
is the corresponding output, i.e., $\mathbf{x}^{\left(j\right)}=f_{\theta_{\text{enc}}^{j}}\left(\mathbf{s}^{\left(j\right)}\right)$,
where $\theta_{\text{enc}}^{j}$ is the encoder parameters of user
$j$. At the AP, the received signal vector at sub-carrier $\alpha$
is 
\begin{equation}
\mathbf{y}_{\alpha}=\sum_{j=1}^{J}\mathbf{h}_{\alpha}^{\left(j\right)}x_{\alpha}^{\left(j\right)}+\mathbf{n}_{\alpha},\label{eq:y_a_UL}
\end{equation}
where $\mathbf{h}_{\alpha}^{\left(j\right)}$ and $\mathbf{n}_{\alpha}$
are the $L\times1$ channel vector from user $j$ to the AP and the
$L\times1$ noise vector of frequency sub-carrier $\alpha$, respectively,
while $x_{\alpha}^{\left(j\right)}$ is the $\alpha$-th entry of
$\mathbf{x}^{\left(j\right)}$, i.e., the transmitted symbol at sub-carrier
$\alpha$ of user $j$, for $j=1,...,J$. For the sake of simplicity
of presentation, we assume that the elements of $\mathbf{h}_{\alpha}^{\left(j\right)}$
and $\mathbf{n}_{\alpha}$ have the same statistics as in Section
II. Our scheme can be straightforwardly extended to the case where
the channels to different users have different variances. For example,
the normalization layer of each user can be scaled by a power allocation
coefficient so that we can allocate more power to users with smaller
channel variances.

Regarding the data decoding at the AP, similar to NC-EA, the combined
energy from $L$ receive antennas is first computed for each sub-carrier,
i.e., $z_{\alpha}=\left\Vert \mathbf{y}_{\alpha}\right\Vert ^{2}$
for $\alpha=1,...,N$. The resulting vector $\mathbf{z}=\left[z_{1},...,z_{N}\right]^{T}$which
collects energy from $J$ users is used as the input of the DNN decoder.
As shown in Fig. \ref{fig:EAMA-UL-struc}, the decoder structure of
the AP consists of $C$ non-linear FC hidden layers with the Tanh
activation, while the output layer is divided into $J$ independent
FC sub-layers of $M$ nodes employing the softmax activation, whose
output is to determine the transmitted data of the corresponding user.
Let us denote $\mathbf{W}_{c}$, $\mathbf{b}_{c}$ and $Q_{c}$ as
the weight, bias and number of nodes, respectively, of the $c$-th
hidden layer of the AP decoder, whose output vector is given by $\mathbf{v}_{c}=\sigma_{\text{Tanh}}\left(\mathbf{W}_{c}\mathbf{v}_{c-1}+\mathbf{b}_{c}\right),$
where $\mathbf{v}_{0}=\mathbf{z}$ and $c=1,...,C.$ As a result,
the output of each final sub-layer can be written by $\mathbf{\hat{s}}^{\left(j\right)}=\sigma_{\text{Softmax}}\left(\mathbf{W}_{C+1}^{\left(j\right)}\mathbf{v}_{C}+\mathbf{b}_{C+1}^{\left(j\right)}\right),$
where $\mathbf{W}_{C+1}^{\left(j\right)}$ and $\mathbf{b}_{C+1}^{\left(j\right)}$
are the weight and bias of the final sub-layer of user $j$, respectively.
Finally, the transmitted message of user $j$ is recovered according
to the largest entry of $\mathbf{\hat{s}}^{\left(j\right)}$.

Note that the structure parameters of uplink NC-EAMA, such as $C$
and $Q_{c}$ $\left(c=1,...,C\right)$, need to be properly selected
based on specific system parameters, such as $N$, $M$ and $J.$
This will be detailed for each experiment in Section IV.

\subsection{Downlink NC-EAMA}

\begin{figure}[tb]
\begin{centering}
\includegraphics[width=1\columnwidth]{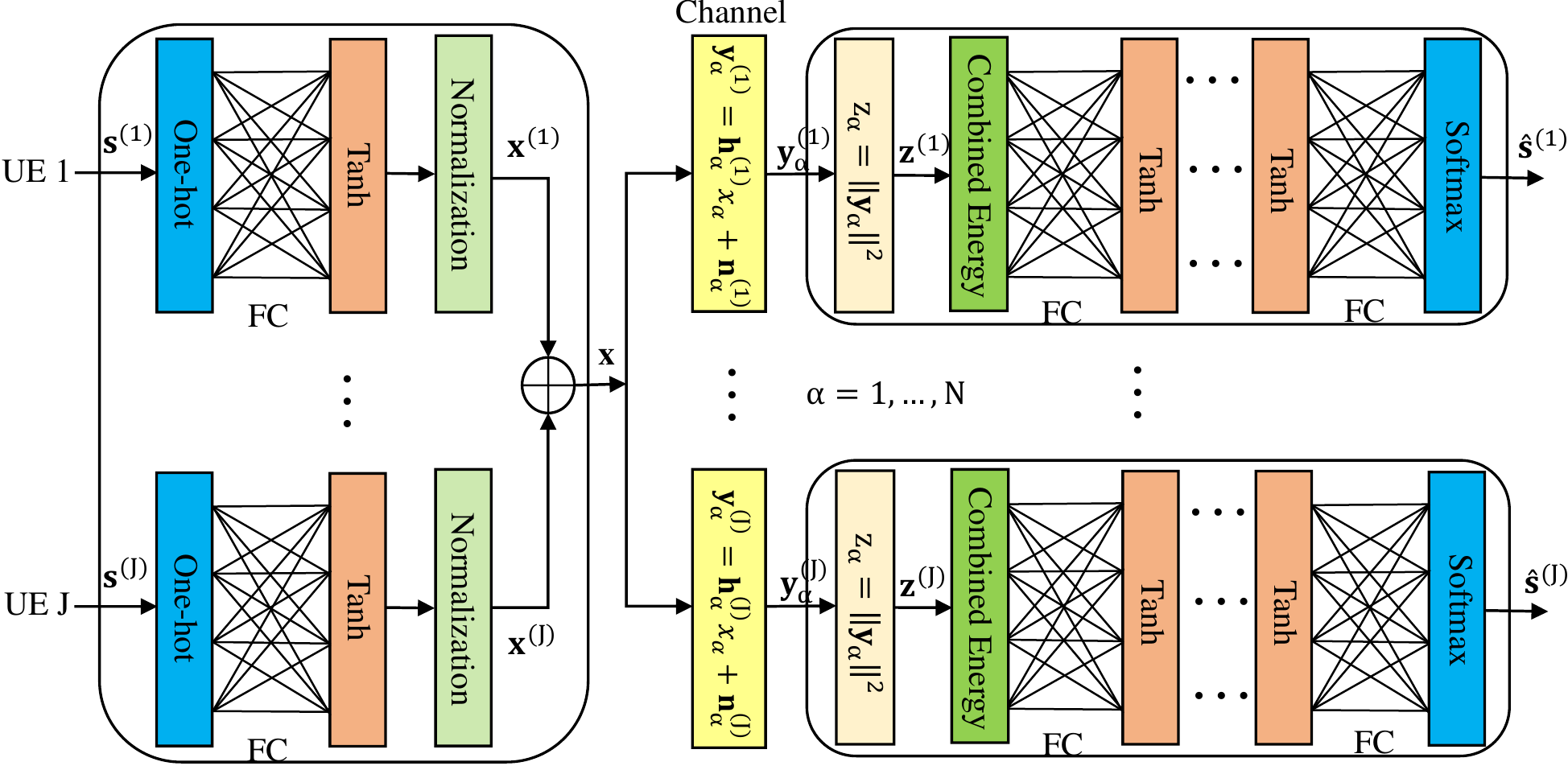} 
\par\end{centering}
\caption{Structure of the downlink NC-EAMA system.\label{fig:EAMA-DL-struc}}
\end{figure}

The DNN structure of downlink NC-EAMA is illustrated in Fig. \ref{fig:EAMA-DL-struc}.
Particularly, the AP equipped with a single antenna communicates simultaneously
with $J$ users, each of which has $L$ antennas, applying the NC-EA
technique on the same $N$ sub-carriers. Note that the users of downlink
NC-EAMA do not need any CSI knowledge to decode their data. The structure
of the AP encoder consists of $J$ sub-networks, whose structure is
the same as the encoder of a single-user NC-EA. Herein, sub-network
$j$ is to encode the data of user $j$ (denoted by the one-hot vector
$\mathbf{s}^{\left(j\right)}$) into the corresponding $N\times1$
codeword $\mathbf{x}^{\left(j\right)}$, for $j=1,...,J.$ Then, the
transmitted vector of the AP that includes the codewords of $J$ users
is determined by $\mathbf{x}=\sum_{j=1}^{J}\mathbf{x}^{\left(j\right)}.$
As such, the AP allocates the same average transmit power for all
users. The received signal vector of user $j$ at sub-carrier $\alpha$
is written by 
\begin{equation}
\mathbf{y}_{\alpha}^{\left(j\right)}=\mathbf{h}_{\alpha}^{\left(j\right)}x_{\alpha}+\mathbf{n}_{\alpha}^{\left(j\right)},\label{eq:y_a_DL}
\end{equation}
where $\mathbf{h}_{\alpha}^{\left(j\right)}$ is the $L\times1$ channel
vector from the AP to the $L$ antennas of user $j$, $\mathbf{n}_{\alpha}^{\left(j\right)}$
is the $L\times1$ noise vector, both have the same statistical models
as presented in the previous section, while $x_{\alpha}$ is the $\alpha$-th
element of $\mathbf{x}$.

The decoder structure of each user in downlink NC-EAMA is similar
to that of single-user NC-EA, except for the fact that it now has
more hidden layers to improve the decoding performance in the presence
of the inter-user interference. In particular, the combined energy
vector $\mathbf{z}^{\left(j\right)}$ of user $j$ is collected as
in \eqref{eq:z_a}, which is then fed to the corresponding DNN decoder.
Denote by $\mathbf{W}_{c}^{\left(j\right)},$ $\mathbf{b}_{c}^{\left(j\right)}$
and $Q_{c}$ the weight, bias and number of nodes of the $c$-th layer
of the decoder of user $j$, for $c=1,...,C+1$, where $C$ denotes
the number of hidden layers. As a result, the output of the decoder
of user $j$ can be expressed by $\mathbf{\hat{s}}^{\left(j\right)}=\sigma_{\text{Softmax}}\left(\mathbf{W}_{C+1}^{\left(j\right)}\sigma_{\text{Tanh}}\left(...\sigma_{\text{Tanh}}\left(\mathbf{W}_{1}^{\left(j\right)}\mathbf{z}^{\left(j\right)}+\mathbf{b}_{1}^{\left(j\right)}\right)\right)+\mathbf{b}_{C+1}^{\left(j\right)}\right),$
which is used to recover the transmitted message of user $j$.

We now highlight some key advantages of the proposed uplink and downlink
NC-EAMA as follows: 
\begin{itemize}
\item The NC-EAMA is highly adaptive and flexible since it can be easily
designed via training for any numbers of users $J$, frequency sub-carriers
$N$ and data streams $M$, as well as any type of the transmission
(downlink or uplink). This flexibility is not available in existing
hand-crafted schemes (e.g., NC-OFDM-IM \cite{NoncoherentOFDMIM},
PAM-MED \cite{PamMed2016}), whose encoder and decoder must be redesigned
in a complicated manner depending on the system requirements. 
\item As a learning scheme, the NC-EAMA allows to jointly optimize both
the transmitter and receiver, which is expected to result in an optimal
performance for each specific system configuration and channel condition,
through properly training the models as shown in the next section. 
\item Compared to NC-OMA schemes, the NC-EAMA can achieve higher diversity
gains since it allows multiple users to utilize all $N$ available
sub-carriers rather than just one or part of $N$ sub-carriers as
in NC-OMA. Thus, our scheme is expected to achieve higher reliability
than NC-OMA, while still enjoying a low decoding complexity when the
decoder requires $C$ and $Q_{c}$ to be small enough. 
\end{itemize}
Note that the aforementioned benefits of the proposed NC-EAMA will
be validated by simulation results in Section IV.

\subsection{Training procedure of NC-EAMA}

The uplink and downlink NC-EAMA schemes are trained offline, using
dataset randomly collected by simulations, based on the known statistics
of the channel and noise vectors. Unlike single-user NC-EA which simply
adopts the MSE loss function for training, we design a new loss function
tailored to NC-EAMA, aiming at fast training convergence to a global
optimum and user fairness regarding the decoding accuracy.

In particular, for brevity, the proposed loss function is written
for each single data sample, as follows: 
\begin{equation}
\mathcal{L}\left(\theta\right)=\sum_{j=1}^{J}\mathcal{E}_{j}+\lambda\sum_{j=1}^{J}\left(\mathcal{E}_{j}-\overline{\mathcal{E}}\right)^{2},\label{eq:Loss-MA}
\end{equation}
where $\mathcal{E}_{j}=\left\Vert \mathbf{s}^{\left(j\right)}-\mathbf{\hat{s}}^{\left(j\right)}\right\Vert ^{2}$
is the least squared error (LSE) of user $j$ and $\overline{\mathcal{E}}=\frac{1}{J}\sum_{j=1}^{J}\mathcal{E}_{j}$,
while $\lambda$ denotes a loss scaling factor. Note that $\lambda$
is an important hyperparameter, which needs to be carefully fine-tuned
while training to ensure a best performance. As seen from \eqref{eq:Loss-MA},
the first term stands for the reconstruction loss, i.e., the total
LSE of all users, while the second term that measures the standard
deviation of the individual LSEs $\mathcal{E}_{j}$ is added to force
them as identical as possible. Interestingly, apart from ensuring
the user fairness as expected, this design enables the DNN models
of NC-EAMA to quickly converge in the training process.

Similar to NC-EA, the SGD-based Adam and Xavier initialization methods
are adopted for training NC-EAMA. Our proposed NC-EAMA models are
trained with multiple training SNRs $\bar{\gamma}_{\text{tr}}$, and
then the trained models are tested with the testing SNRs $\bar{\gamma}_{\text{te}}$
being the same as $\bar{\gamma}_{\text{tr}}$ in order to yield the
best performance. The details of selecting other training parameters,
such as epochs, batch size, learning rate, training and testing data
sizes, and particularly the loss scaling factor $\lambda$, will be
provided for various experiments in the next section.

\section{Simulation results}

We carry out extensive simulations to verify the error performance
of the proposed NC-EA and uplink/downlink NC-EAMA schemes in comparison
with baseline schemes. Particularly, the performance of our schemes
is evaluated in terms of the block error rate (BLER) versus the average
SNR per bit $E_{b}/N_{0}$, where $E_{b}=mE_{s}/N$ denotes the average
transmit power per bit.\footnote{Since the bit error rate (BER) analysis delivers the same message
as the BLER analysis, for the sake of simplicity, we include the BLER
results only.} A block error event occurs when a message of $m$ bits of each user
transmitted over a block of $N$ sub-carriers is incorrectly decoded.
We also present a decoding complexity comparison at the end of this
section.

\subsection{BLER Performance of NC-EA}

\begin{table}
\caption{NC-EA training parameters \label{tab:NC-EA-train-para}}

\centering{}%
\begin{tabular}{|c|c|}
\hline 
Parameters  & Values\tabularnewline
\hline 
\hline 
Epoch  & $10^{3}$\tabularnewline
\hline 
Batch size  & 128\tabularnewline
\hline 
Train size  & $2\times10^{4}$\tabularnewline
\hline 
Test size  & $10^{6}$\tabularnewline
\hline 
Learning rate  & 0.001\tabularnewline
\hline 
$Q$  & $16,32,64$ for $M=4,8,16$\tabularnewline
\hline 
\end{tabular}
\end{table}

We consider NC-OFDM-IM \cite{NoncoherentOFDMIM} and PAM-MED \cite{PamMed2016}
that use the noncoherent ML detector \eqref{eq:ML}, as baselines
of the proposed NC-EA. In particular, NC-OFDM-IM only operates at
low data rates of $<1$ bps/Hz, while PAM-MED operates at higher data
rates of $\ge1$ bps/Hz. Note that since PAM-MED is designed for single-carrier
transmission only, we independently employ it on each sub-carrier
for comparison with our multicarrier scheme. Similar to NC-EA, the
noncoherent ML detector of these baselines also needs to know the
average SNR for data decoding. The configurations of NC-EA, NC-OFDM-IM
and PAM-MED are denoted by $(N,M)$, $(N,K)$ and $(N,D)$, respectively,
where we recall that $N$ is the number of sub-carriers, $M=2^{m}$
with $m$ being the size of the transmitted message of NC-EA, $K$
is the number of active sub-carriers of NC-OFDM-IM, and $D$ is the
modulation order of PAM-MED. The training parameters of NC-EA are
given in Table \ref{tab:NC-EA-train-para}.

\begin{figure}[tb]
\begin{centering}
\includegraphics[width=0.93\columnwidth]{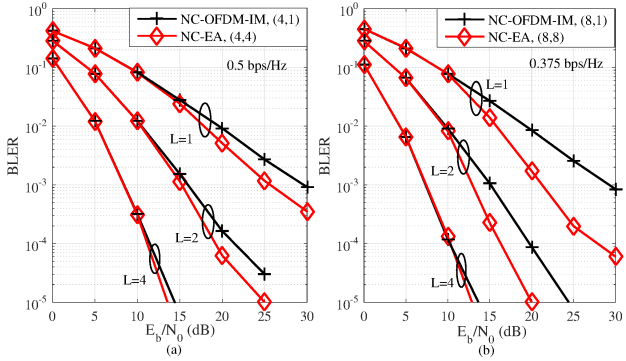} 
\par\end{centering}
\caption{BLER comparison between the proposed NC-EA and NC-OFDM-IM \cite{NoncoherentOFDMIM}
when (a) $(N,M)=(4,4)$ and (b) $(N,M)=(8,8)$. \label{fig:EA01}}
\end{figure}

Fig. \ref{fig:EA01} compares the BLER performance between the proposed
NC-EA and NC-OFDM-IM when (a) $(N,M)=(4,4)$ and (b) $(N,M)=(8,8)$,
and $L=1,2,4$, at the data rates of 0.5 and 0.375 bps/Hz, respectively.
Herein, NC-OFDM-IM activates $K=1$ sub-carrier in both cases to achieve
the same data rates as NC-EA. It is shown via Fig. \ref{fig:EA01}
that our scheme outperforms the baseline, especially at high SNRs
and small $L$. For example, in Fig. \ref{fig:EA01}(b), at the BLER
of $10^{-3}$, NC-EA achieves 8 dB and 2 dB SNR gains over the baseline
when $L=1$ and $2$, respectively. This indicates that the unitary
codewords of NC-OFDM-IM are not an optimal design for every SNR level,
while the proposed NC-EA can learn to return the optimal codewords
for any SNR levels via training.

\begin{figure}[tb]
\begin{centering}
\includegraphics[width=0.92\columnwidth]{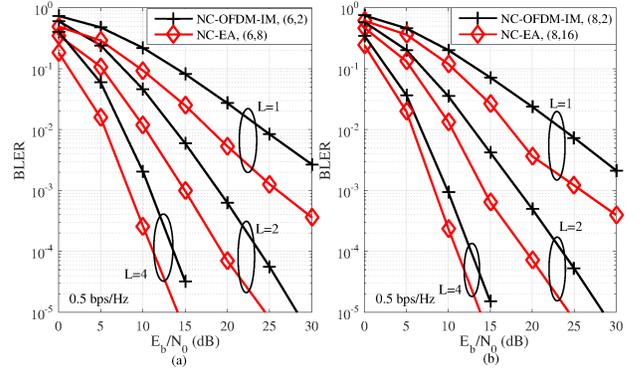} 
\par\end{centering}
\caption{BLER comparison between the proposed NC-EA and NC-OFDM-IM \cite{NoncoherentOFDMIM}
when (a) $(N,M)=(6,8)$ and (b) $(N,M)=(8,16)$. \label{fig:EA02}}
\end{figure}

In Fig. \ref{fig:EA02}, we illustrate the BLER comparison between
the proposed NC-EA and NC-OFDM-IM at higher data rates with $M>N$
and $K>1$, particularly when (a) $(N,M)=(6,8)$ and (b) $(N,M)=(8,16)$,
and $K=2$. Note that for given $N$, NC-OFDM-IM needs $K>1$ to support
higher data rate transmissions. Unlike the previous figure, it is
observed from Fig. \ref{fig:EA02} that NC-EA performs much better
than the baseline in whole SNR regions, even when $L$ increases.
For example, in Fig. \ref{fig:EA02}(a), at a BLER of $10^{-2}$,
our scheme provides about 6, 3 and 2 dB SNR gains over the baseline
when $L=1,2$ and $4$, respectively. This improvement comes from
the fact that when $(N,K)=(6,2)$, the baseline has a total of $C\left(6,2\right)=15$
possible unitary codewords, however, it only utilizes 8 codewords
to convey 3 bits, which is obviously an inefficient and sub-optimal
design. By contrast, the proposed EA approach which can learn to optimize
codewords appears to ideally address the drawback of the hand-designed
baseline.

\begin{figure}[tb]
\begin{centering}
\includegraphics[width=0.8\columnwidth]{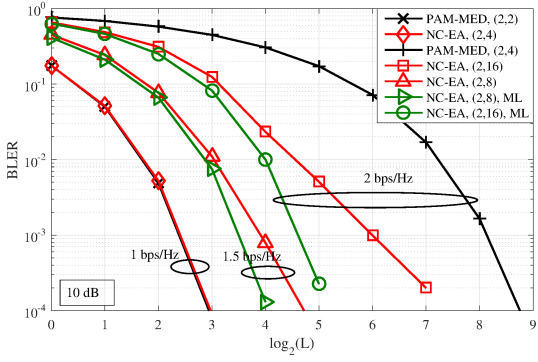} 
\par\end{centering}
\caption{BLER comparison between the proposed NC-EA and PAM-MED \cite{PamMed2016}
when $E_{b}/N_{0}=10$ dB, $N=2$, $M=\{4,8,16\}$ and $D=2,4$. \label{fig:EA03}}
\end{figure}

Fig. \ref{fig:EA03} depicts the BLER performance versus $\log_{2}(L)$
of the proposed NC-EA and PAM-MED at the data rates of $\ge1$ bps/Hz,
when $E_{b}/N_{0}=10$ dB, $N=2$, $M=4,8,16$, $D=2,4$ and $L=1,2,...,2^{9}$.
Note that the baseline scheme employs the PAM-MED technique with $D=2$
and $4$ on each sub-carrier to support 1 and 2 bps/Hz data rates.
We point out NC-OFDM-IM is not considered since it is not able to
work at more than $1$ bps/Hz. The performance of NC-EA with the noncoherent
ML decoder is included, besides that of the DNN decoder. As seen in
Fig. \ref{fig:EA03}, at 1 bps/Hz, NC-EA and PAM-MED have the same
performance since when $D=2$, PAM-MED becomes an on-off keying (OOK)
scheme, which is known to be optimal in this case. At higher data
rate, i.e., 2 bps/Hz, NC-EA considerably outperforms PAM-MED. For
instance, at 2 bps/Hz, our scheme needs less than 32 antennas to achieve
a BLER of $10^{-2}$, while PAM-MED requires more than 128 antennas.
Moreover, while the baseline is unable to support the data rate of
1.5 bps/Hz, our scheme still performs well with the BLER curve lying
between the ones of 1 and 2 bps/Hz. This clearly confirms the advantage
in terms of higher flexibility of NC-EA compared to hand-crafted baselines,
such as NC-OFDM-IM and PAM-based schemes. Finally, in NC-EA, the optimal
ML outperforms the DNN decoder as expected, especially when $M$ and
$L$ get larger, at the cost of substantial computational complexity.

\begin{figure}[tb]
\begin{centering}
\includegraphics[width=0.93\columnwidth]{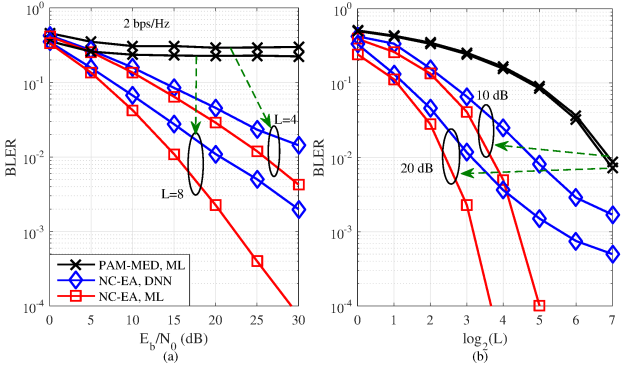} 
\par\end{centering}
\caption{BLER comparison between the single-carrier NC-EA and PAM-MED \cite{PamMed2016}
when $(N,M)=(1,4)$, (a) $L=1,4,8$ and (b) $E_{b}/N_{0}=10$, $20$
dB. The NC-EA employs ML and DNN decoders, while PAM-MED uses the
ML decoder. \label{fig:EA04}}
\end{figure}

Note that the proposed NC-EA works well not only for multicarrier
but also for single-carrier transmissions as shown in Fig. \ref{fig:EA04}.
Particularly, Fig. \ref{fig:EA04} illustrates the BLER comparison
between single-carrier NC-EA and PAM-MED when $(N,M)=(1,4)$, wherein
Fig. \ref{fig:EA04}(a) presents the BLER versus $E_{b}/N_{0}$ when
$L=1,4,8$, while Fig. \ref{fig:EA04}(b) depicts the BLER versus
$\log_{2}(L)$ when $E_{b}/N_{0}=10$ and $20$ dB. Here, NC-EA employs
both the DNN and noncoherent ML decoders. As seen in Fig. \ref{fig:EA04}(a),
while the baseline suffers from a prohibitive error floor, our scheme
achieves much better BLER, which decreases with increasing the SNR.
Moreover, the ML decoder significantly enhances the performance of
NC-EA compared to the DNN decoder which is known to be a sub-optimal
decoder. As observed in Fig. \ref{fig:EA04}(b), our scheme again
outperforms PAM-MED. For example, at 20 dB, NC-EA with either DNN
or ML decoder only needs about 8 antennas to achieve the BLER of $10^{-2}$,
while the baseline requires 128 antennas.

\subsection{BLER Performance of NC-EAMA}

We note that NC-EAMA is the first multicarrier NC-NOMA scheme that
allows multiple users to share the same set of frequency sub-carriers.
Thus, it is reasonable to compare NC-EAMA with NC-OMA schemes which
are based on either noncoherent IM or PAM techniques. In particular,
each user in NC-OMA is evenly allocated $n=N/J$ sub-carriers to independently
employ NC-OFDM-IM and PAM-MED for low and high data rate transmissions,
respectively. The training parameters of NC-EAMA are the same as that
of NC-EA in Table \ref{tab:NC-EA-train-para}, except for the number
of hidden layers of the decoder $C$ is fixed to 2 in all experiments,
and the corresponding hidden nodes denoted by $\left\{ Q_{1},Q_{2}\right\} $
will be provided for each specific experiment. Especially, the loss
scaling factor $\lambda$ is empirically selected from the set of
$\mathcal{F}=\left\{ 0,1,5,10,20\right\} .$ Note from our experiments
that there does not exist an optimal $\lambda$ for any system parameters
$(J,N,M)$ as well as SNR levels.

\begin{figure}[tb]
\begin{centering}
\includegraphics[width=0.93\columnwidth]{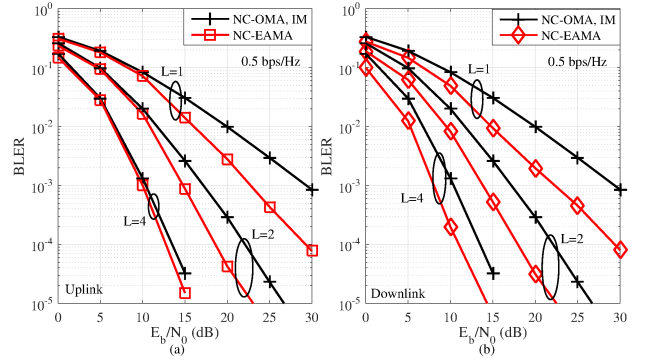} 
\par\end{centering}
\caption{BLER comparison between the (a) uplink and (b) downlink NC-EAMA, and
NC-OMA with IM, when $(J,N,M)=(2,4,2)$ and $L=1,2,4$. \label{fig:MA01}}
\end{figure}

Fig. \ref{fig:MA01} compares the BLER performance between the proposed
(a) uplink and (b) downlink NC-EAMA, and NC-OMA with IM, when $(J,N,M)=(2,4,2)$
and $L=1,2,4$. Herein, NC-EAMA employs $\{8,16\}$ and $\{4,8\}$
hidden nodes for the uplink and downlink decoders, respectively. It
is shown via Fig. \ref{fig:MA01} that NC-EAMA has better BLER than
NC-OMA for both uplink and downlink, especially when $L$ is small
and the SNR gets larger. For example, at the BLER of $10^{-3}$, the
uplink NC-EAMA achieves an SNR gain of 6.5 dB and 2.5 dB over NC-OMA
when $L=1$ and $2$, respectively. The SNR gain achieved by the downlink
NC-EAMA is even larger. In fact, NC-EAMA allows multiple users to
simultaneously spread their transmit powers across all $N$ sub-carriers
in an optimized manner via training, achieving higher diversity gains
than NC-OMA, whose users employ $n\ll N$ sub-carriers only.

\begin{figure}[tb]
\begin{centering}
\includegraphics[width=0.8\columnwidth]{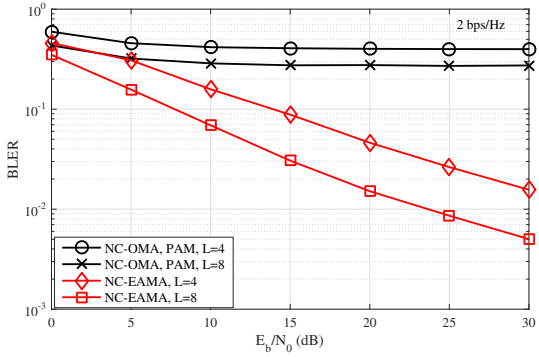} 
\par\end{centering}
\caption{BLER comparison between the uplink NC-EAMA and NC-OMA with PAM, at
a data rate of 2 bps/Hz, when $(J,N,M)=(2,2,4)$ and $L=4,8$. \label{fig:MA02}}
\end{figure}

Fig. \ref{fig:MA02} depicts the BLER comparison between uplink NC-EAMA
and NC-OMA with PAM at higher data rate, i.e., 2 bps/Hz, when $(J,N,M)=(2,2,4)$
and $L=4,8$. The decoder of NC-EAMA has $\left\{ 16,32\right\} $
hidden nodes. Note that NC-OMA with IM does not work at the considered
high data rate. We can see from Fig. \ref{fig:MA02} that the BLER
of the proposed scheme decreases with increasing the SNR, thus is
much better than that of the baseline which incurs a very high error
floor. The same observation can also be made for the downlink transmission
that we omit for the sake of brevity.

\begin{figure}[tb]
\begin{centering}
\includegraphics[width=0.8\columnwidth]{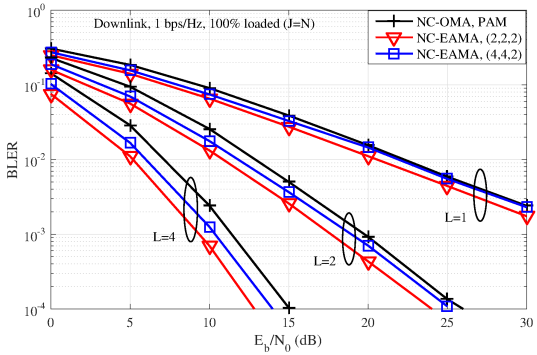} 
\par\end{centering}
\caption{BLER comparison between the downlink NC-EAMA and NC-OMA with PAM when
$(J,N,M)=(2,2,2)$, $(4,4,2),$ and $L=1,2,4$. \label{fig:MA03}}
\end{figure}

In Fig. \ref{fig:MA03}, we compare the performance of downlink NC-EAMA
and NC-OMA with PAM when $(J,N,M)=(2,2,2)$, $(4,4,2),$ and $L=1,2,4$,
i.e., the system is 100\% fully-loaded with $J=N$. In this case,
each user of NC-OMA employs the single-carrier PAM-MED, i.e., OOK
transmission, while each user of NC-EAMA has $\left\{ 4,8\right\} $
hidden nodes in the decoder. It is interesting from Fig. \ref{fig:MA03}
that NC-EAMA outperforms the baseline for all SNR values, in particular
the performance gap between them is larger when $L$ and $N$ increase.
This is due to the fact that more sub-carriers used for each user
leads to more diversity gains achieved by NC-EAMA over NC-OAM whose
users use only one sub-carrier. For uplink, we found in our experiments
that NC-EAMA performs similarly to NC-OMA with PAM in the same setting.

\begin{figure}[tb]
\begin{centering}
\includegraphics[width=0.8\columnwidth]{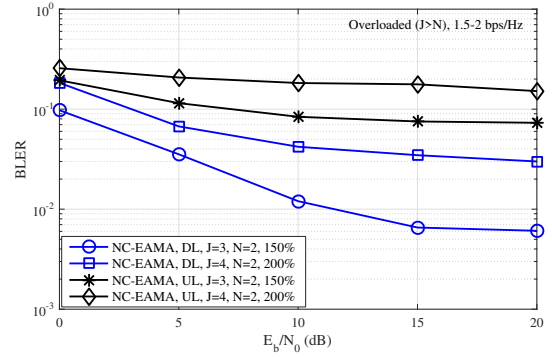} 
\par\end{centering}
\caption{BLER performance of the uplink and downlink NC-EAMA under overloaded
transmissions, when $J=3,4$, $N=2$, $M=2$ and $L=8.$ \label{fig:MA04}}
\end{figure}

\begin{figure}[tb]
\begin{centering}
\includegraphics[width=0.8\columnwidth]{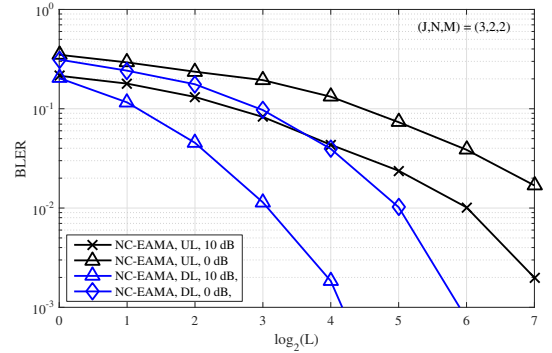} 
\par\end{centering}
\caption{BLER versus $\log_{2}\left(L\right)$ of the NC-EAMA under overloaded
transmissions, when $(J,N,M)=(3,2,2)$, $E_{b}/N_{0}=0$ dB and $10$
dB. \label{fig:MA05}}
\end{figure}

Fig. \ref{fig:MA04} presents the BLER of uplink and downlink NC-EAMA
under overloaded transmissions with $J>N$, when $J=3,4$, $N=2$,
$M=2$ and $L=8$. Herein, our schemes employ $\{16,32\}$ and $\{8,16\}$
hidden nodes for uplink and downlink decoders, respectively. It should
be noted that hand-crafted NC-OMA schemes are unable to support overloaded
transmissions, while our schemes perform relatively well under 150\%
and 200\% overloading, especially in the downlink, as shown in Fig.
\ref{fig:MA04}. However, due to the severe inter-user interference
when the number of users increases, while the number of sub-carriers
is limited, the BLER of NC-EAMA experiences an error floor at increasing
SNRs. Thus, it is essential to enable NC-EAMA to support more users
under limited sub-carrier resources, while still ensuring a good performance.
One solution is to increase the number of antennas $L$ as shown in
Fig. \ref{fig:MA05}, where the reliability of NC-EAMA is noticeably
enhanced as $L$ gets larger, even at a small SNR of 0 dB. Also, an
noncoherent ML decoder can be derived for NC-EAMA to improve its performance
compared to the DNN decoder, which is considered as our future work.

\subsection{Complexity Comparison}

\begin{table}
\caption{Runtimes of NC-EA and IM/PAM baselines in microseconds \label{tab:com-nc-ea}}

\centering{}%
\begin{tabular}{|c|c|c|c|}
\hline 
$(N,M)$  & NC-EA  & IM \cite{NoncoherentOFDMIM}  & PAM \cite{PamMed2016}\tabularnewline
\hline 
\hline 
$(8,8)$  & 8.15  & 4.72  & N/A\tabularnewline
\hline 
$(6,8)$  & 7.89  & 16.23  & N/A\tabularnewline
\hline 
$(8,16)$  & 9.62  & 26.85  & N/A\tabularnewline
\hline 
$(1,4)$  & 6.57/3.54  & N/A  & 3.56\tabularnewline
\hline 
\end{tabular}
\end{table}

We investigate the detection complexity of the proposed schemes in
comparison with baseline schemes using the noncoherent ML detector.
In particular, we measure the runtime of successfully decoding a transmitted
message of $m=\log_{2}(M)$ bits at the receiver by considering that
all schemes are implemented on MATLAB of the same machine for fairness.
Note that the trained models of our proposed schemes on Tensorflow
are converted into MATLAB to compute the runtimes. Since the effect
of $L$ on the decoding complexity of all schemes is the same at the
step of computing the combined energy, we simply adopt $L=4$ to measure
the runtimes for simplicity.

Table \ref{tab:com-nc-ea} compares the runtimes in microseconds $\left(\mu s\right)$
between NC-EA and NC-OFDM-IM or PAM-MED baselines (abbreviated as
IM/PAM on the table). Here, the system parameters $(N,M)$ on Table
\ref{tab:com-nc-ea} are associated with some of figures in Subsection
IV.A. It is shown via Table \ref{tab:com-nc-ea} that NC-EA requires
runtimes comparable to the baselines, which are only a few microseconds.
Particularly, compared to NC-OFDM-IM, the runtime of NC-EA is larger
when $N=M=8$ and is smaller when $N<M$. This is because when $N<M$,
NC-OFDM-IM needs to activate more sub-carriers, i.e., $K>1$, which
significantly increases its detection complexity compared to the case
of $N=M$, i.e., $K=1$. By contrast, the complexity of NC-EA does
not increase much when $N$ or $M$ increases due to its simple decoder
structure with a few hidden nodes as shown in Table \ref{tab:NC-EA-train-para}.
This clearly confirms the benefits of our proposed scheme over NC-OFDM-IM
in terms of the receiver complexity as presented in Section II. Moreover,
compared to PAM-MED when $(N,M)=(1,4)$, NC-EA with the DNN decoder
requires a longer runtime, i.e., $6.57$ $\mu s$. However, using
the ML decoder, our scheme demands a runtime similar to the baseline
with around $3.5$ $\mu s$, while achieving much better BLER as in
Fig. \ref{fig:EA04}. Note that as $N,\,M$ get larger the DNN may
have lower complexity than ML decoder. For example, when $(N,M)=(16,64)$,
the runtime of the DNN decoder with 128 hidden nodes is 18 $\mu s$,
which is much lower than that of the ML with 294 $\mu s$. 

\begin{table}
\caption{Runtimes of NC-EAMA and NC-OMA baselines in microseconds \label{tab:com-nc-eama}}

\centering{}%
\begin{tabular}{|c|c|c|c|}
\hline 
$(J,N,M)$  & NC-EAMA  & NC-OMA/IM  & NC-OAM/PAM \tabularnewline
\hline 
\hline 
$(2,4,2),$ UL  & 6.03  & 4.17  & N/A\tabularnewline
\hline 
$(2,4,2),$ DL  & 7.91  & 4.17  & N/A\tabularnewline
\hline 
$(2,2,4),$ UL  & 6.41  & N/A  & 3.45\tabularnewline
\hline 
$(4,4,2),$ DL  & 7.74  & N/A  & 3.28\tabularnewline
\hline 
\end{tabular}
\end{table}

The runtimes of NC-EAMA and NC-OMA baselines based on either IM or
PAM are depicted in Table \ref{tab:com-nc-eama}, where the system
parameters $(J,N,M)$ are associated with the figures in Subsection
IV.B. Similar to NC-EA, both the proposed uplink and downlink NC-EAMA
schemes demand slow runtimes in decoding data, which are only several
microseconds. Compared to the NC-OMA baselines, NC-EAMA requires larger
runtimes. This is understandable since the baselines only need to
employ the single-carrier detection independently, while our scheme
has to detect the signals across all sub-carriers.

Finally, to better understand the impact of system parameters, please
refer to Table \ref{tab:com-bigO} which illustrates the Big-O complexity
of the noncoherent ML and proposed DNN decoders, where both downlink
and uplink NC-EAMA require two hidden layers, while the ML decoder
\eqref{eq:ML} is used for all baselines above. Here, the term $\mathcal{O}\left(4NL\right)$
which appears in all detection schemes refers to the complexity of
computing the received energy from $L$ receive antennas as in \eqref{eq:z_a}.
It is shown via Table \ref{tab:com-bigO} that the decoding complexities
of the proposed schemes increase with the numbers of nodes in the
hidden layers of the DNN decoders. However, when these numbers are
not too large, our schemes yield comparable or even lower complexity
compared with the ML decoder of the baselines as shown in the previous
runtime comparison.

\begin{table}
\caption{Complexity of noncoherent ML and proposed DNN decoders \label{tab:com-bigO}}

\centering{}%
\begin{tabular}{|c|c|}
\hline 
Detection schemes  & Complexity\tabularnewline
\hline 
\hline 
Noncoherent ML  & $\mathcal{O}\left(10NM\right)+\mathcal{O}\left(4NL\right)$\tabularnewline
\hline 
NC-EA  & $\mathcal{O}\left(NQ+QM\right)+\mathcal{O}\left(4NL\right)$\tabularnewline
\hline 
Uplink NC-EAMA  & $\mathcal{O}\left(NQ_{1}+Q_{1}Q_{2}+JQ_{2}M\right)+\mathcal{O}\left(4NL\right)$\tabularnewline
\hline 
Downlink NC-EAMA  & $\mathcal{O}\left(NQ_{1}+Q_{1}Q_{2}+Q_{2}M\right)+\mathcal{O}\left(4NL\right)$\tabularnewline
\hline 
\end{tabular}
\end{table}

\section{Conclusion}

We have explored the potential of DL in noncoherent energy-based systems
under fading channels, which do not involve any CSI estimation, for
both single-user and multi-user transmissions under the multicarrier
SIMO framework. In particular, it was shown that the proposed single-user
NC-EA can provide a range of advantages over existing hand-crafted
schemes, such as higher reliability, higher SE and higher flexibility
with comparable or lower detection complexity. Interestingly, the
NC-EA still performs well even with single-carrier transmissions.
For multiuser scenarios, the proposed NC-EAMA based on the multicarrier
MU-SIMO framework is also highly flexible as it can be designed to
accommodate any number of users, sub-carriers, antennas and data streams,
as well as any transmission directions, while current hand-crafted
schemes are unable to enjoy such highly flexible designs. More importantly,
developing the opportunities of the NC-NOMA scheme, NC-EAMA can achieve
much higher reliability than NC-OMA counterparts, while still enjoying
a low decoding complexity. We showed that the proposed NC-EAMA still
works well even with overloaded transmissions, especially when the
number of antennas is large enough. Hence, our proposed schemes are
appropriate for MTCs which demand reliability, low latency and low
complexity.

 \bibliographystyle{IEEEtran}
\phantomsection\addcontentsline{toc}{section}{\refname}\bibliography{Ref}

\end{document}